\documentclass[english, 5p, number, sort&compress 
]{elsarticle}
\usepackage[T1]{fontenc}
\usepackage[latin1]{inputenc}
\usepackage{graphicx}
\usepackage{amssymb}
\usepackage{verbatim}
\usepackage{epic,eepic}
\usepackage{babel}
\DeclareMathAlphabet{\mathpzc}{OT1}{pzc}{m}{it} 
\usepackage{lmodern}
\usepackage{color}
\begin{document}

\title{Experimental determination of the quasi-projectile mass with measured neutrons}

\author[Gan]{P.~Marini\corref{cor}} \ead{marini@ganil.fr}
\author[Chem,Tamu]{A.~Zarrella}
\author[Tamu,LNS]{A.~Bonasera}
\author[Chem,Tamu]{G.~Bonasera}
\author[Chem,Tamu]{P.~Cammarata}
\author[Chem,Tamu]{L.~Heilborn}
\author[Chem,Tamu]{Z.~Kohley\fnref{altaff}}
\author[Tamu]{J.~Mabiala}
\author[Chem,Tamu]{L.~W.~May}   
\author[Tamu]{A.~B.~McIntosh}
\author[Chem,Tamu]{A.~Raphelt}
\author[Tamu,greece]{G.~A.~Souliotis}
\author[Chem,Tamu]{S.~J.~Yennello}

\cortext[cor]{Corresponding author}
\fntext[altaff]{Present address: National Superconducting Cyclotron Laboratory, Michigan State University, East Lansing, Michigan 48824, USA.}

\address[Gan]{GANIL, Bd. H. Becquerel, BP 55027 - 14076 CAEN, France}
\address[Chem]{Chemistry Department, Texas A\&M University, College Station, TX-77843, USA}
\address[Tamu]{Cyclotron Institute, Texas A\&M University, College Station, TX-77843, USA}
\address[LNS]{Laboratori Nazionali del Sud, INFN, via Santa Sofia, 62, 95123 Catania, Italy}
\address[greece]{Laboratory of Physical Chemistry, Department of Chemistry, National and Kapodistrian University of Athens, 15771 Athens, Greece}

\begin{abstract}
The investigation of the isospin dependence of multifragmentation reactions relies on precise reconstruction of the fragmenting source. The criteria used to assign free emitted neutrons, detected with the TAMU Neutron Ball, to the quasi-projectile source are investigated in the framework of two different simulation codes. Overall and source-specific detection efficiencies for multifragmentation events are found to be model independent. The equivalence of the two different methods used to assign experimentally detected charged particles and neutrons to the emitting source is shown. The method used experimentally to determine quasi-projectile emitted free neutron multiplicity is found to be reasonably accurate and sufficiently precise as to allow for the study of well-defined quasi-projectile sources.
\end{abstract}

\begin{keyword}
Isospin physics, quasi-projectile mass reconstruction, multifragmentation, neutron detection
\end{keyword}

\maketitle


\section{Introduction}
The study of multifragmentation as a function of isospin has recently become an area of interest \cite{xu2000, veselsky2000,dempsey96,yennello94, yang99, winchester2000, martin2000, laforest99, baran2005,tsang2001_2, souliotis04,wuenschel2010,marini2011} because of its connection with the isospin-dependent part of the equation of state \cite{baran2005prc}. 
Much of the early experimental work was performed by varying the isospin content, $N/Z$, of the beam and/or target and measuring observables as a function of the $N/Z$ of the reacting system.
Later, several works showed that the key parameter in understanding isospin dynamics might be the isospin content of the fragmenting system rather than that of the reacting system \cite{veselsky2000,rowland03,wuenschel2010,marini2011}. 
For instance, the excited quasi-projectile (QP, the momentarly existing primary fragment, remnant of the projectile after a non-central collision) produced in peripheral and semi-peripheral collisions at intermediate energies can access  a rather broad distribution in $N/Z$  and  the centroid of the distribution is shifted toward the valley of stability \cite{veselsky2000,rowland03,wuenschel2010}. 
More specifically, the importance of  charge and mass characterization of the fragmenting source was pointed out in several works \cite{wuenschel2009,marini2011} investigating the symmetry energy term of the nuclear equation of state. Isoscaling \cite{tsang2001_2} and m-scaling \cite{huang2010Mscaling} have been shown to significantly improve when selecting  quasi-projectiles within a narrow window of $N/Z$ \cite{wuenschel2009, marini2011}.  Similarly, an agreement among different methods of extracting the symmetry energy coefficient was reported in \cite{marini2011}, provided that stringent constraints were placed on the isotopic asymmetry of the QP source.

The abundant emission of charged particles and neutrons accompanying these reactions often tend to obscure the primary processes, making it difficult to characterize intermediate systems formed in the collision and  study the underlying reaction mechanisms.
Different approaches were used to determine the mean $N/Z$ of the QP source. One approach uses isotopically resolved charged fragments to reconstruct the QP on an event-by-event basis and relies on models to include the evaporated neutrons in the  measured $N/Z$ of the QP \cite{rowland03,rowland_thesis, keksis2010}. A second approach, which relies on the hypothesis of early fragment formation \cite{marie98, shetty03}, uses fragment yield ratios of pairs of isobars to determine the average $N/Z$ of the QP \cite{keksis_thesis, keksis2010}.

The availability of  neutron calorimeters coupled to $4\pi$ charged particle detectors allows for direct event-by-event measurement of free emitted neutrons and charged particles. This makes it possible to perform an event-by-event  QP charge and mass measurements, independent of models.

In this work we will present and explore the reliability of the criteria that we use to experimentally assign detected free neutrons to the QP fragmenting source. 
For this purpose, overall and source-specific neutron detection efficiencies of the Neutron Ball detector will be calculated using the HIPSE-SIMON simulation code \cite{lacroix04,durand92} for reaction systems corresponding to the experimental data described in Sec. \ref{sec:section2}. Neutrons will be assigned to the QP using both a source tag provided by the HIPSE code and a velocity selection experimentally used for charged particles (see Sec. \ref{sec:qp_rec}). The results obtained from both methods of QP neutron assignment and their comparison will be reported throughout this paper to show their equivalence. In addition results obtained with the Constrained Molecular Dynamics model \cite{papa2001} for the same reaction systems will be presented alongside HIPSE-SIMON results to illustrate the model independence of our criteria. Finally, the free neutron assignment method will be applied to our experimental data and the QP neutron distributions will be shown for three different systems.

This article is organized as follows. In Sec. \ref{sec:section2} we present the experimental setup and the QP reconstruction procedure using information on the emitted free neutrons. In Sec. \ref{sec:analysis} the calculation of the overall, QP and QT efficiencies for multifragmentation events is presented, as obtained by analysing two different models. Moreover, the reliability of the QP reconstruction as well as the effect of the experimental  geometrical and detection efficiencies of the Neutron Ball are analysed. An application of the presented criteria to experimental data is presented in Sec. \ref{sec:exp}. Conclusions are presented in Sec. \ref{sec:conclusions}.

\section{Experimental conditions and procedures}\label{sec:section2}
Several experiments were recently carried out at the Texas A$\&$M University on the K500 superconducting Cyclotron to investigate the isospin-dependent part of the equation of state by looking at QP fragmentation events. Such studies rely on the availability of an experimental setup which provides event-by-event information on the emitted charged particles and neutrons, and on a QP reconstruction procedure capable of identifying the fragmenting QP in charge, mass and exitation energy. In the following we will describe the experimental setup and the QP reconstruction procedure used in these experiments.

\subsection{The  experimental setup}\label{sec:Experimental_setup}
The $4\pi$ NIMROD-ISiS array \cite{wuenschelNimrod, wuenschelNeutronBall} is used for the detection of charged particles and neutrons. 
The detector telescopes, arranged on $15$ rings centered on the beam axis, are composed of one or two silicon detectors backed by  a CsI(Tl) crystal with PMT readout. 
Isotopic resolution, essential for isospin physics analyses, is obtained up to $Z=17$. Details on the mass identification resolution and procedures can be found in \cite{zach_thesis, kohley2012}.
The  charged-particle array is housed inside the TAMU  Neutron Ball \cite{wuenschelNeutronBall}, which measures the free neutron multiplicity.

The Neutron Ball consists of $6$ scintillator tanks, filled with a pseudocumene based liquid scintillator (NE224) loaded with $0.3\,$wt. $\%$ Gd which surrounds the vacuum chamber.
Four large quadrant-shaped detectors are sandwiched between the two end cups  tanks, creating a cylinder.
Each scintillator tank is read by $3$ or $4$ fast photomultiplier tubes (quadrants and emispheres, respectively). 
Following a nuclear reaction, prompt $\gamma$ rays and neutrons dissipate energy in the scintillaor, producing a flash. Relatively low energy neutrons tend to be thermalized in the scintillator and diffuse through the detector for a period of microseconds before undergoing capture reactions mainly with Gd and H nuclei. After
 the fast flash PMT signals, a delayed  $100\,\mu s$ gate is opened. The number of delayed flashes in the gate is proportional to the number of neutrons emitted in the triggering reaction. Following this gate, a second $100\,\mu s$ gate is opened. The multiplicity measured in the second gate provides information on the background. The beam is prevented from entering the detector during the counting gates by a deflector. The background measurement also takes into account  the possible activation of material in the chamber. 

Further details on the Neutron Ball geometry and electronics, the detection mechanisms and the detector intrinsic efficiency can be found in \cite{schmitt95, wuenschelNimrod}.

\subsection{Quasi-projectile reconstruction procedure}\label{sec:qp_rec}

Particle and event selections are performed to select quasi-projectile (QP) fragmentation events.

The quasi-projectile source is reconstructed on an event-by-event basis by associating charged particle fragments with the QP source based on a velocity selection in which each particle's longitudinal velocity ($v_{z}$) is compared to the $v_{z}$ of the largest fragment.  For $Z = 1$ fragments, $v_{z}$ must be within $\pm 65\%$ of the largest fragment's $v_{z}$.  For $Z=2$ and   $Z\geq 3$, the cut is at $\pm 60\%$ and $\pm 45\%$, respectively.
This cut, later referred to as \textit{Vcut}, is intended to remove fragments from non-projectile-like sources and can be applied only to measured charged particles since the Neutron Ball  does not measure neutron velocities. 

In order to select on QP fragmentation events, the total charge of the  fragments detected in the $4\pi$ array and included in the reconstruction was chosen to be between $83\%$ and $100\%$ of the beam charge
 (\textit{SumZ}). Finally, limits are placed on the deformation of the source, as measured by the quadrupole momentum, to select a class of events that are, on average, spherical \cite{wuenschel2010}. The quadrupole momentum, calculated from the measured particle momenta in the quasi-projectile frame, $\sum_{i} p_{\|_{i}}^{2}/\sum_{i} \frac{1}{2} p_{\bot_{i}}^{2}$, is required to be between $0.4$ and $2.7$. 
 This will be later referred to as \textit{Qcut}.

Free neutrons measured by the Neutron Ball were used to calculate the number of free neutrons emitted by the quasi-projectile
($N_{QP}$) using the relation \cite{wuenschel2010}:
\begin{equation}\label{eq.nqp}
N_{QP} = \frac{N_{det}} {\varepsilon_{QP}-\frac{N_{T}}{N_{P}} \varepsilon_{QT}}.
\end{equation}
where $N_{det}$ is the measured neutron multiplicity.
In order to calculate the number of free neutrons emitted by the QP, the total multiplicity is  corrected by the efficiency for detecting free neutrons emitted from the quasi-projectile ($\varepsilon_{QP}$) and the quasi-target($\varepsilon_{QT}$) for our reactions.  Equation \ref{eq.nqp} assumes that QP and QT have the same relative $N/Z$ as the projectile and target and that the main contributions to neutron emission come from QP and QT emissions, while other sources (such as, for instance,  pre-equilibrium emission) are neglected. $N_{P}$ and $N_{T}$ are the projectile and target neutron numbers, respectively. The determination of the efficiencies for the QP and QT sources will be presented in Sec. \ref{subsec:overallQPQTeff}.
Through this formulation, the number of neutrons emitted from the QP can be calculated from only a total multiplicity of free neutrons.

\section{Analysis}\label{sec:analysis}
Recently, experimental data on $^{64}$Zn+$^{64}$Zn, $^{70}$Zn+$^{70}$Zn and $^{64}$Ni+$^{64}$Ni at $35\,$AMeV were used to investigate the isospin dependence of the caloric curve \cite{mcintosh12} and to compare different methods to extract infomation on the isospin-dependent part of the equation of state \cite{marini2011}. Both studies pointed out the need for placing stringent constraints on the isotopic asymmetry of the analysed QP source. The uncertainty in the QP identification arises mainly from the assignment of detected free neutrons to the QP source. In order to develop a   procedure for assigning free emitted neutrons to the QP source,  a systematic study was performed on the predictions of two different models, HIPSE-SIMON \cite{lacroix04,durand92} and CoMD \cite{papa2001}, for the $^{64}$Zn+$^{64}$Zn, $^{70}$Zn+$^{70}$Zn and $^{64}$Ni+$^{64}$Ni reactions.  The number of events generated with the HIPSE-SIMON code and the CoMD code is $1600000$ and $40400$, respectively.

The HIPSE \cite{lacroix04} model is a phenomenological event generator dedicated to the description of nuclear collisions in the intermediate energy range. The approach is mainly based on the geometrical participant-spectator picture and accounts for both statistical and dynamical effects. The after-burner phase is achieved using the SIMON event generator \cite{durand92} which takes into account all possible decay channels from neutron evaporation to symmetric fission. Information on the source of the final particle is provided in the output of the calculation which makes the code very suitable to investigate the source-specific efficiency of the Neutron Ball.

The CoMD model \cite{papa2001} was also  used, and the results were compared to the  HIPSE-SIMON results. CoMD  is a constrained molecular dynamics model for a fermionic system, suitable for the study of heavy colliding systems. The time evolution of each particle was followed, in our calculation, up to $3000\,$fm/c, to assure a full de-excitation of the fragment. No information on the particle-emitting source is provided by the code.

An experimental filter was implemented to simulate the Neutron Ball response.
The Neutron Ball efficiency was calculated in Ref.\cite{wada04} using the GCALOR code coupled to the GEANT-3 simulation package to simulate the Neutron Ball incorporated into NIMROD. Indeed, the large amount of material inside the Neutron Ball chamber, due to the presence of the charged particle array, can result in the  scattering, absorption and generation of neutrons. The calculations were done for neutrons emitted isotropically from the target with energies from $5$ to $200$ MeV (in $5\,$MeV steps). An efficiency of $0.6$ was obtained for all neutron energies averaged over all angles. However, in  heavy ion reactions around Fermi energy, higher energy neutrons are significantly peaked in the forward direction. Moreover, the charged particle array has more material at forward angles and the Neutron Ball is not symmetric with respect to the target. Therefore, the neutron efficiency was calculated as a function of both neutron energy and polar angle. Further details can be found in Ref. \cite{wada04}. This energy and angle dependent efficiency was implemented in the experimental  filter.

The experimental filter and the QP reconstruction procedure (i.e. \textit{VCut, ZCut} and \textit{QCut}) were applied to the simulated data. The reduced impact parameter distribution of fully equilibrated quasi-projectile fragmentation events ranges from $0.5$ to $0.9$ and it is peaked around $0.7$ for the CoMD calculation. For HIPSE results, the distribution is peaked around $0.6$ and extends from $0.5$ to slightly above $0.8$. For this reason  we will constrain our analysis to $0.5\leq b_{red} \leq 0.9$ for the rest of this paper. Data were divided into $4$ bins, according to the reduced impact parameter: $b_{red1}=0.55\pm0.05$, $b_{red2}=0.65\pm0.05$, $b_{red3}=0.75\pm0.05$, $b_{red4}=0.85\pm0.05$. 
HIPSE-SIMON results are not  shown for bins which contain fewer than $100$ events.

\subsection{Overall, QP and QT efficiencies determination}\label{subsec:overallQPQTeff}
The efficiency of the Neutron Ball as a function of energy and angle from the target position for isotropic emission is well established \cite{wada04}. However, we are interested in reconstructing QPs resulting from multifragmentation events. Therefore it is important to know the effective neutron efficiencies for these types of events.

\subsubsection{Overall efficiency}
 The overall Neutron Ball efficiency was determined as:
\begin{equation}
\varepsilon_{tot} = \frac{N_{det}}{N_{raw}}
\end{equation}
where $N_{raw}$ is the number of  neutrons predicted by the simulations and $N_{det}$ is the number of neutrons passing the experimental filter, i.e. the detected neutrons.
The calculation was performed  on both HIPSE-SIMON and CoMD simulated data.
\begin{figure}
\centering
\includegraphics[width=0.95\columnwidth]{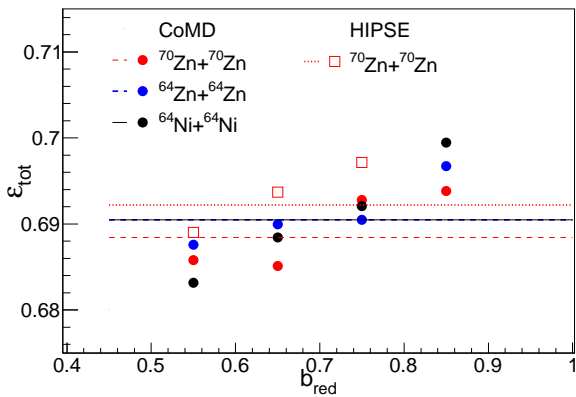} 
\caption{(Color online) Overall efficiencies determined from CoMD and HIPSE simulated data. Lines correspond to  overall efficiencies obtained without sorting the data by reduced impact parameter. Lines corresponding to $^{64}$Zn+$^{64}$Zn and $^{64}$Ni+$^{64}$Ni overlap.}
\label{fig:TotalEfficiency}
\end{figure}

The overall efficiencies, $\varepsilon_{tot}$, are shown as horizontal lines in Fig. \ref{fig:TotalEfficiency}. The dotted line corresponds to values obtained by HIPSE-SIMON calculation for the $^{70}$Zn+$^{70}$Zn reaction, while the other lines correspond to values obtained by CoMD simulations for the three different reactions. The calculated uncertainty in the efficiencies reported throughout the rest of this text are of the order of or below $1\%$ and will be omitted in the following.
Efficiencies of about $69\%$ are obtained for CoMD simulated data for the three reactions.
The values obtained for the $^{64}$Zn and $^{64}$Ni reactions  overlap.
No significant dependence on the analysed reaction is observed, therefore only values obtained for the $^{70}$Zn+$^{70}$Zn reaction will be shown in the remainder  of the figures in the paper.  A slightly higher $\varepsilon_{tot}$ of $69.2\%$ is obtained from the HIPSE-SIMON simulation (dotted line) for the $^{70}$Zn+$^{70}$Zn reaction, which is   within $0.3\%$ of that obtained from the CoMD results.

Overall efficiencies, sorted by reduced impact parameter, are also plotted for the three analysed reactions in Fig. \ref{fig:TotalEfficiency}. Full and open symbols correspond to values obtained from CoMD and HIPSE-SIMON results, respectively.  No HIPSE simulated data are available in the highest $b_{red}$ bin, due to lack of statistics. 
Although we observe a slight dependence of the total efficiency on the reduced impact parameter, the obtained values are in overall agreement to within less than $1\%$. As for the global efficiencies, there is no dependence on the reaction and the values obtained from CoMD and HIPSE agree to within $1\%$. 

\subsubsection{Efficacy of the velocity selection}\label{sec:VCut effect}
The Neutron Ball efficiency for quasi-projectile emitted neutrons was determined as
\begin{equation}\label{eq:qp_eff}
\varepsilon_{QPn} = \frac{N_{QPn,\,det}}{N_{QPn,\,raw}},
\end{equation}
where $N_{QPn,\,raw}$ is the number of QP-neutrons  predicted by the simulation and $N_{QPn,\,det}$ is the number of the detected QP-emitted neutrons, i.e. those neutrons emitted from the QP which pass the experimental filter. QT efficiencies can be obtained from Eq. \ref{eq:qp_eff} by considering QT-emitted neutrons rather than QP.
In the HIPSE-SIMON output, neutrons can be assigned to the QP based on the source tag provided by the model. Also, a velocity constraint, similar to that applied to charged particles, can be applied to assign neutrons to the QP. The efficacy of such a constraint is investigated by comparing the efficiencies for QP-emitted neutrons obtained using the two selections.
This is important for the following analysis because, in CoMD results, information on the emitting source (i.e. the source tag) is not available.

As discussed in Sec. \ref{sec:qp_rec}, the velocity selection allows us to select QP-emitted charged particles based on their longitudinal velocity with respect to the largest fragment emitted in the event (projectile-like fragment, PLF).
From simulation rersults, the selection limits of the neutron longitudinal velocity relative to the largest fragment were determined by correcting for the Coulomb boost that acts on the protons. The QP reference frame was reconstructed by taking into account all the charged particles accepted by the \textit{VCut}. Neutrons are not considered at this point since, experimentally, neutron velocities are not measured and, therefore, cannot be included in the QP velocity ($v_{QP}$) determination.
The Coulomb boost, $E_{C}$, in the quasi-projectile reference frame was determined as the difference between the proton and neutron average kinetic energies
\begin{equation}
\langle E_{n}^{(QP)} \rangle = \langle E_{p}^{(QP)} \rangle - E_{C}
\end{equation}
or
\begin{equation}\label{eq:CoulombCorrection}
\frac{1}{2}m_{n}v_{n}^{(QP)^{2}} = \frac{1}{2}m_{p}v_{p}^{(QP)^{2}}- E_{C}.\\
\end{equation}
where $v_{n}^{(QP)}$ and $v_{p}^{(QP)}$ are the neutron and proton velocities in the QP reference frame, respectively.
Next, we bring Eq. \ref{eq:CoulombCorrection} into the lab reference frame and divide by the PLF velocity ($v_{PLF}$).  We substitute the used upper and lower limits for the quantity $v_{p}^{(lab)}/v_{PLF}^{(lab)}$ ($1.65$ and $0.35$, see Sec. \ref{sec:qp_rec}) and assume that the QP velocity is similar to the largest fragment velocity ($v_{QP}\sim v_{PLF}$).  This leads to the conclusion that the longitudinal neutron velocity relative to the largest fragment limits are $\pm49\%$.
The calculation was performed for $^{64}$Zn+$^{64}$Zn, $^{70}$Zn+$^{70}$Zn and $^{64}$Ni+$^{64}$Ni reactions, sorting the data by reduced impact parameter. The 
values obtained for the three reactions agree to within about $2\%$. \\

It is important to ensure that the the source tag and the velocity selection for associating free neutrons with the QP produce similar results. For this we investigate the percentages of HIPSE-tagged QP neutrons which also pass the \textit{Vcut}. The obtained values are reported for each $b_{red}$ bin in Tab. \ref{tab.qp_np_that_make_VCut} for the $^{70}$Zn+$^{70}$Zn reaction.
More than $90\%$ of QP-tagged neutrons pass the velocity selection for $b_{red}>0.6$, and about $85\%$ are accepted for the most central considered $b_{red}$ range.
A similar analysis was performed on QP-tagged protons and the obtained values are also reported in the table. A very good agreement is observed between the values obtained for neutrons and protons. In both cases the velocity selection is very effective  in accepting QP-emitted neutrons and protons, and the selection only gets better for the more peripheral collisions.\\

\begin{table}
\centering
\begin{tabular}{|c|c|c|c|c|}
\hline
 $b_{red}$ & $0.5-0.6$  & $0.6-0.7$ & $0.7-0.8$ & $0.8-0.9$ \\ 
\hline  
 neutrons & $85.5\%$ & $93.2\%$ & $96.7\%$ & $98.5\%$  \\ 
\hline
 protons & $82.6\%$ & $90.3\%$ & $95.4\%$ & $98.6\%$  \\ 
\hline 
\end{tabular} 
\caption{ Percentages of QP-tagged neutrons and protons accepted by the velocity selection for each reduced impact parameter bin for the $^{70}$Zn+$^{70}$Zn reaction, as predicted by HIPSE-SIMON simulation.}
\label{tab.qp_np_that_make_VCut}
\end{table}

\begin{figure}
\centering
\includegraphics[width=0.95\columnwidth]{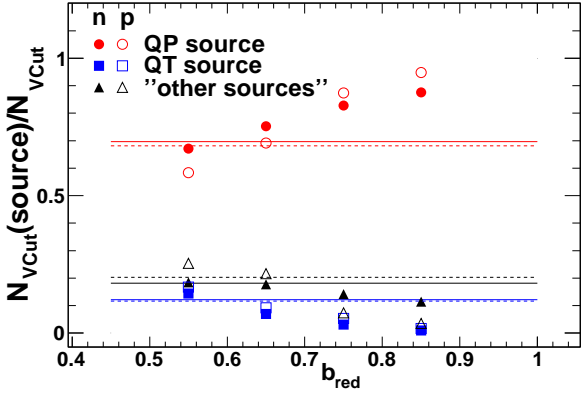} 
\caption{(Color online) Fraction of \textit{VCut}-accepted neutrons (full symbols) and protons (open symbols) produced by each source relative to the total number of neutrons that pass the \textit{VCut}, according to HIPSE-SIMON calculation. Full and dashed lines are the values obtained without sorting the data by $b_{red}$, for neutrons and protons respectively.}
\label{fig:HIPSEWhichSourceForVCutAcceptedNP}
\end{figure}

The capability of the \textit{VCut} to select only QP-emitted neutrons (and protons) is investigated by looking at the emitting source of  particles accepted by the velocity selection (i.e.  \textit{VCut}-accepted particles). Sources will be classified as QP, QT and ``other sources'', the latter including  emitting sources other than the QP and QT.
\begin{table}[b]
\centering
\begin{tabular}{|c|c|c|c|}
\hline
                       &\multicolumn{3}{c|}{Source}\\
\cline{2-4}
                                    & QP        & QT & Others \\
\hline
$N_{VCut}^{n}(source)/N_{VCut}^{n}$ &  $70\%$   & $12\%$ & $18\%$  \\
$N_{VCut}^{p}(source)/N_{VCut}^{p}$ &  $68\%$   & $12\%$ & $20\%$ \\
\hline 
\end{tabular} 
\caption{Percentage of \textit{VCut}-accepted neutrons and protons produced by each source relative to the total number of neutrons passing the \textit{VCut} for the $^{70}$Zn+$^{70}$Zn reaction, as predicted by HIPSE-SIMON simulation.} 
\label{tab.which_source_for_Vaccepted_Np}
\end{table}
Table \ref{tab.which_source_for_Vaccepted_Np} shows the emitting source, according to HIPSE, of \textit{VCut}-accepted neutrons and protons  for the $^{70}$Zn+$^{70}$Zn reaction. 
\textit{VCut}-accepted neutrons (and protons) mainly ($\approx70\%$) come from the QP source. Quasi-target and ``other source'' contributions are of the order of $\approx10\%$ and $\lessapprox20\%$, respectively. These values are plotted in Fig. \ref{fig:HIPSEWhichSourceForVCutAcceptedNP} as full (emitted neutrons) and dashed (emitted protons) red, blue and black lines for QP/QT/``other source''. Values sorted by $b_{red}$ are also plotted in Fig. \ref{fig:HIPSEWhichSourceForVCutAcceptedNP} as circles, squares and triangles for QP, QT and ``other sources''. Open and full symbols correspond to protons and neutrons, respectively. 
The fraction of \textit{VCut}-accepted neutrons belonging to the QP source (red circles) increases as the $b_{red}$  increases while the contamination from neutrons emitted by the QT and ``other sources'' decreases, consequently. Similar results are obtained for the other reactions. The velocity selection appears to be very effective in selecting QP-neutrons throughout the range of $b_{red}$ and gets increasingly better for more peripheral collisions.

\begin{figure}
\centering
\includegraphics[width=0.95\columnwidth]{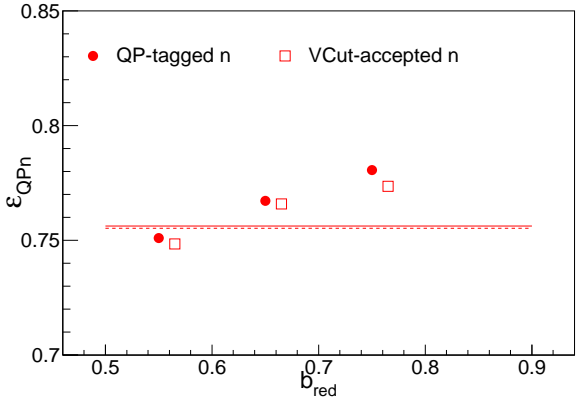} 
\caption{Quasi-projectile neutron efficiencies calculated according to eq.\ref{eq:qp_eff} using QP-tagged ($\varepsilon_{QP(tagged)}$) and \textit{VCut}-accepted ($\varepsilon_{QP(Vcut)}$) neutrons from HIPSE-SIMON results for $^{70}$Zn+$^{70}$Zn reaction. Full and dashed lines are the values obtained using all $b_{red}$ (see text).}
\label{fig:HIPSEQPEfficiencyTagVsVCut}
\end{figure}

Quasi-projectile efficiencies were calculated according to Eq. \ref{eq:qp_eff}, using both QP-tagged and \textit{VCut}-accepted neutrons from HIPSE-SIMON.
  The obtained $\varepsilon_{QP_{n}}$, integrated over the $0.5-0.9$ reduced impact parameter range, are reported in Tab. \ref{tab:hipse_QPeff_lab_and_Vcut} for all  reactions as $\varepsilon_{QP(tagged)}$  and  $\varepsilon_{QP(Vcut-acc.)}$, respectively. The obtained values are in good agreement  and no dependence on the reaction is observed.
   In Fig. \ref{fig:HIPSEQPEfficiencyTagVsVCut}, QP neutron efficiencies are plotted for the $^{70}$Zn+$^{70}$Zn reaction as a function of the reduced impact parameter as determined by QP-tagged neutrons (circles) and by velocity-accepted neutrons (squares). The average quasi-projectile neutron efficiencies are also plotted for QP-tagged neutrons (full line) and velocity-accepted neutrons (dashed line). A small dependence on the impact parameter is observed. Nevertheless, the obtained values are in agreement. Such agreement indicates that the velocity selection is an effective way to identify neutrons emitted from the quasi-projectile source.

From an experimental point of view, the velocity selection cannot be applied to neutrons.  In order to reconstruct a QP, then, the charged particles and neutrons must be assigned to the QP differently - by the \textit{VCut} for charged particles and by Eq. \ref{eq.nqp} for neutrons.
In the simulation, our newly derived neutron velocity selection criteria behaves very similarly to the proton velocity cut that has been previously used to reconstruct QP's and, therefore, suggests that it is a successful procedure for assigning neutrons to the QP.  Furthermore, this neutron velocity selection is consistent with assignment of neutrons to the QP based on the source tag from HIPSE. We conclude, then, that, within the results of our simulations, the method of QP neutron assignment based on the source tag selection method, which is unavailable for CoMD results,  is an effective substitute for the velocity cut selection.
\begin{table}
\centering
\begin{tabular}{|c|c|c|c|}
\hline
                    &   & HIPSE-SIMON                                & CoMD\\
                    \hline
$^{70}$Zn+$^{70}$Zn & $\varepsilon_{QP(tagged)}$    & $75.7\%$      & $-$\\
                    & $\varepsilon_{QP(Vcut-acc.)}$ & $75.5\%$      &  $76.7\%$ \\
                    & $\varepsilon_{QT}$ & $61.8\%$      & $60.8\%$ \\
                    \hline
$^{64}$Zn+$^{64}$Zn & $\varepsilon_{QP(tagged)}$    & $75.6\%$      &$-$\\
                    & $\varepsilon_{QP(Vcut-acc.)}$ & $75.3\%$     & $76.6\%$ \\
                    & $\varepsilon_{QT}$ & $61.1\%$      & $61.2\%$ \\
\hline
$^{64}$Ni+$^{64}$Ni & $\varepsilon_{QP(tagged)}$    & $76.0\%$      &$-$\\
                    & $\varepsilon_{QP(Vcut-acc.)}$ & $75.6\%$      & $77.5\%$ \\
                    & $\varepsilon_{QT}$ & $63.0\%$      & $60.4\%$ \\
\hline 
\end{tabular} 
\caption{Quasi-projectile ($\varepsilon_{QP}$) and quasi-target ($\varepsilon_{QT}$) efficiencies calculated according to Eq. \ref{eq:qp_eff}  for both  models. Tagged and \textit{VCut}-accepted neutrons were used to determine $\varepsilon_{QP(tagged)}$ and $\varepsilon_{QP(Vcut-acc.)}$, respectively (see text).}
\label{tab:hipse_QPeff_lab_and_Vcut}
\end{table}


\subsubsection{Quasi-projectile and quasi-target efficiencies}

Quasi-projectile and quasi-target efficiencies were calculated according to Eq. \ref{eq:qp_eff} by considering QP and QT-tagged neutrons from the HIPSE simulation and by considering \textit{VCut}-accepted neutrons from CoMD simulation. 
The number of QP-emitted neutrons predicted by the CoMD simulation ($N_{QPn,\,raw}$) was determined on the basis of the longitudinal neutron momentum. Indeed, the distribution of neutron momenta  presents two peaks that are clearly separated, especially for the most peripheral $b_{red}$. These peaks correspond to neutrons emitted from the QP (forward in center-of-mass) and the QT (backward in center-of-mass) sources. The QP/QT efficiencies are reported in Tab. \ref{tab:hipse_QPeff_lab_and_Vcut} for all the reactions and for both the models. Values of about $76\%$ and $61\%$ are obtained for $\varepsilon_{QP}$ and $\varepsilon_{QT}$, respectively.  No dependence on the system is observed. Moreover, the values obtained for HIPSE-SIMON and CoMD simulated data are in agreement within less than $2\%$. This suggests that these values are largely model independent.

\begin{figure}
\centering
\includegraphics[width=0.95\columnwidth]{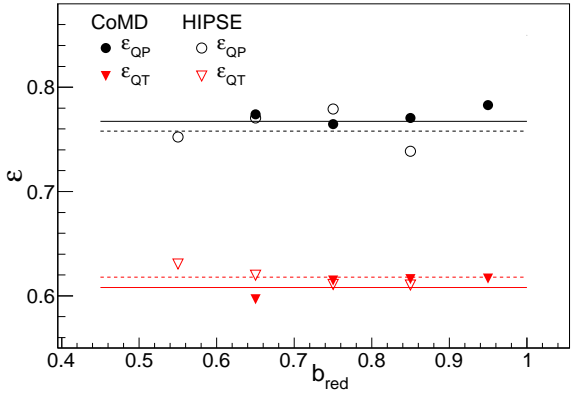} 
\caption{(Color online) Quasi-projectile (circles) and quasi-target (triangles) emitted neutron detection efficiencies obtained for the $^{70}$Zn+$^{70}$Zn reaction from CoMD (full symbols and line) and HIPSE-SIMON (open symbols and dashed line) simulated data.}
\label{fig:QPQTEfficiencies}
\end{figure}

In Fig.\ref{fig:QPQTEfficiencies}, $\varepsilon_{QP}$ (black lines) and $\varepsilon_{QT}$ (red lines) are plotted for the $^{70}$Zn+$^{70}$Zn reaction  for CoMD (full lines) and HIPSE (dashed lines) simulations. The efficiencies for neutrons sorted by impact parameter are also plotted in Fig.\ref{fig:QPQTEfficiencies}.  Full and empty symbols indicate values obtained for CoMD and HIPSE-SIMON simulated data, respectively. The efficiency value obtained from HIPSE results for the most peripheral collisions ($b_{red}=0.8-0.9$) is affected by a larger uncertainty due to lack of statistics. The CoMD-extracted efficiency value for $b_{red}=0.5-0.6$ is not presented because for the more central  collisions the distribution of neutron momenta does not present two separated peaks. No significant dependence on $b_{red}$ is observed and there is good agreement between the values obtained from CoMD and HIPSE-SIMON simulations.

\subsection{ An event-by-event analysis of the Neutron Ball}

\begin{figure}
\centering
\includegraphics[width=0.95\columnwidth]
{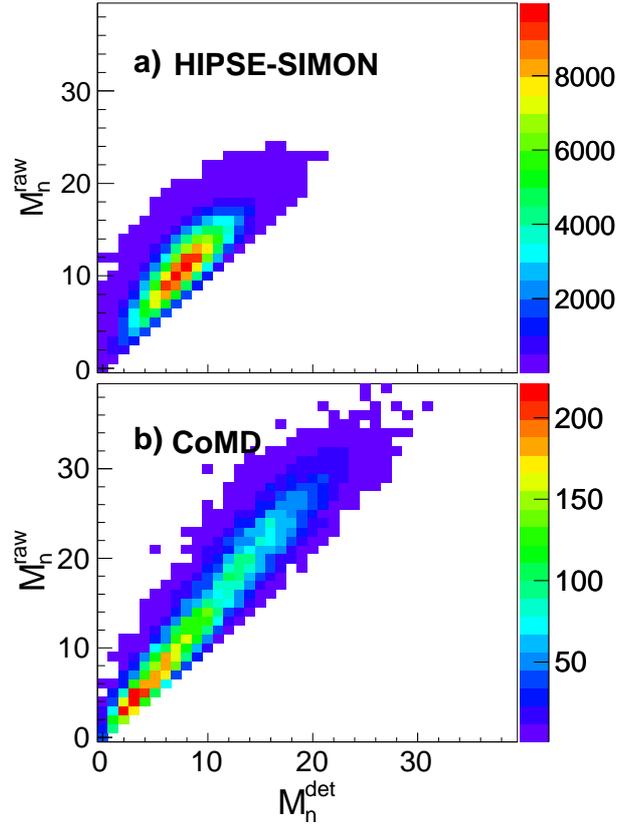} 
\caption{(Color online) Multiplicity of emitted free neutrons ($M_{n}^{raw}$) versus multiplicity of detected free neutrons ($M_{n}^{det}$) as obtained from HIPSE-SIMON (a) and CoMD (b) simulations for the $^{70}$Zn+$^{70}$Zn reaction (see text).}
\label{fig:filterEffTotalN}
\end{figure}

Once  the overall, QP and QT free neutron detection efficiencies for multifragmentation events were determined, we  investigated the event-by-event response of the Neutron Ball.
In Fig. \ref{fig:filterEffTotalN} the multiplicity of  free neutrons for each event ($M_{n}^{raw}$) is plotted versus the multiplicity of detected free neutrons ($M_{n}^{det}$), i.e. the number of neutrons that pass the experimental filter. Figure \ref{fig:filterEffTotalN}(a) and \ref{fig:filterEffTotalN}(b) were obtained by analyzing the HIPSE-SIMON and CoMD simulated data, respectively, for the $^{70}$Zn+$^{70}$Zn reaction. The two models predict, for the same reaction, very different numbers of emitted neutrons. Nevertheless a rather narrow correlation of $M_{n}^{raw}$ and $M_{n}^{det}$ is observed in both cases. The standard deviation, $\sigma$, of the  total multiplicity $M_{n}^{raw}$ is approximately constant for each $M_{n}^{det}$ value for both models. The weighted average $\sigma$ is $1.92$
and it varies from $1.5$ to $2.0$ for $0\leq M_{n}^{det}\leq 17$ for HIPSE results. For CoMD simulated data the average $\sigma$ is 
 $2.5$ and it varies from $1.2$ to $3$ for $0\leq M_{n}^{det}\leq 26$. The number of detected neutrons, $M_{n}^{det}$, is smaller than the number of raw neutrons, consistent with a detection efficiency smaller than $1$. The observed correlation shows that information  on the real free neutron multiplicity can be extracted from the experimentally measured neutron number.

\begin{figure}
\centering
\includegraphics[width=0.95\columnwidth]
{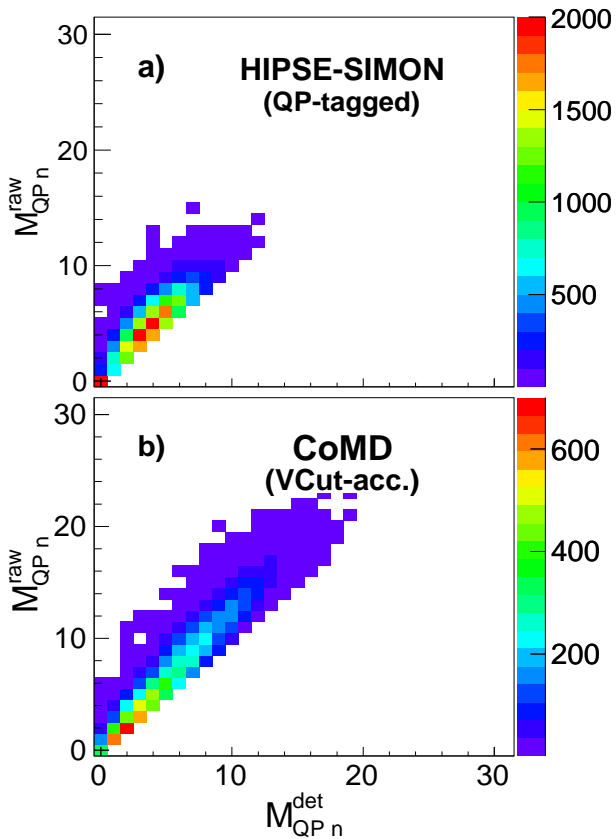} 
\caption{(Color online) Multiplicity of QP-emitted free neutrons, $M_{n}^{raw}$ (unfiltered), versus multiplicity of detected QP-emitted free neutrons, $M_{n}^{det}$ (filtered), as obtained from HIPSE-SIMON (a) and CoMD (b) simulations for the $^{70}$Zn+$^{70}$Zn reaction. Neutrons are assigned to the QP source according to the tag and the \textit{VCut} for HIPSE-SIMON  and CoMD  calculations, respectively (see text).}
\label{fig:filterEffQPN}
\end{figure}

The effect of the experimental filter on the number of QP-emitted neutrons was also investigated. In Fig. \ref{fig:filterEffQPN}   the total number of QP-emitted neutrons is plotted versus the number of detected QP-emitted neutrons. Figure \ref{fig:filterEffQPN}(a) is obtained for HIPSE simulated data and QP-tagged neutrons are considered. The narrow correlation observed for the raw neutron multiplicity and the multiplicity of detected neutrons (Fig. \ref{fig:filterEffTotalN}) is observed also between the raw QP-emitted neutron multiplicity ($M_{QP_{n}}^{raw}$) and the multiplicity of detected neutrons emitted by the QP ($M_{QP_{n}}^{det}$). In this case the width is slightly smaller, with an average value of  
$\langle \sigma \rangle =1.22$, and does not vary significantly with $M_{QP_{n}}^{det}$.

Figure \ref{fig:filterEffQPN}(b) is obtained for CoMD simulated data. Neutrons are assigned to the QP source according to the velocity selection described in Sec. \ref{sec:VCut effect}. Once more, the correlation between $M_{QP_{n}}^{raw}$ and $M_{QP_{n}}^{det}$ is observed, with a constant average width of 
 $2.19$ neutrons. From these observations we can conclude that the selection of the neutron emission source does not wash out the strong correlation that we see in both panels of Fig. \ref{fig:filterEffTotalN}.  The Neutron Ball response remains sharply peaked which suggests reliable event-by-event neutron multiplicity measurements. Moreover, it speaks to the quality of the velocity selection that, when applied to CoMD results, the correlation does not broaden.
Similar narrow correlations are observed for QP-emitted protons and for each of the other reactions. 
This gives us confidence that, with a proper treatment of the efficiencies, we are able to determine the  number of neutrons to  assign to the QP.

\subsection{Accuracy and precision of the QP neutron assignment method}\label{sec:how good is our formula}

\begin{figure}[h!]
\centering
\includegraphics[width=0.95\columnwidth]
{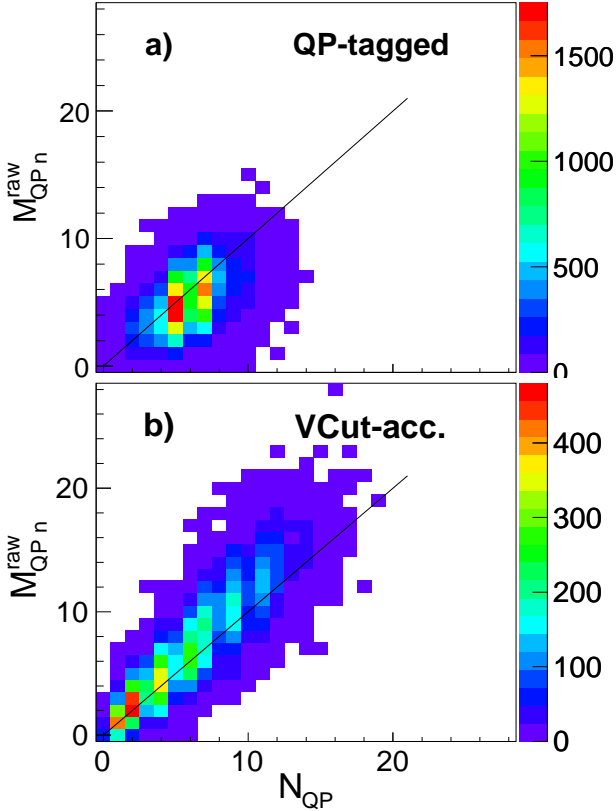}
\caption{(Color online) Multiplicity of QP-emitted neutrons ($M_{QP_{n}}^{raw}$) versus multiplicity  of free neutron assigned to the reconstructed QP, $N_{QP}$, according to Eq.\ref{eq.nqp}, for the $^{70}$Zn+$^{70}$Zn. (a. HIPSE-SIMON and b. CoMD calculations)}
\label{fig:howgoodOurFormulaIs}
\end{figure}

Knowing the source-specific efficiencies for multifragmentation events, as determined from HIPSE or CoMD simulations,  allows one to calculate the number of QP-emitted neutrons from the total number of detected free neutrons using Eq. \ref{eq.nqp}. We recall that $\varepsilon_{QP}$ and $\varepsilon_{QT}$ were found to be model independent in our analysis. The total number of detected neutrons ($M_{n}^{det}$) predicted by HIPSE-SIMON and CoMD calculations were used, event-by-event, to calculate the number of neutrons to be assigned to the QP ($N_{QP}$). In Fig.\ref{fig:howgoodOurFormulaIs} the raw number of QP-emitted neutrons ($M_{QP_{n}}^{raw}$) is plotted against the multiplicity of QP-assigned neutrons as predicted by HIPSE-SIMON (a) and CoMD (b) simulations. The full line indicates $N_{QP} = M_{QP_{n}}^{raw}$. Each $N_{QP}$ value corresponds to a distribution of raw QP-emitted neutrons   predicted by the simulation. 
In order to better depict the correlations in Fig. \ref{fig:howgoodOurFormulaIs} we can obtain a distribution of raw QP-emitted neutrons for each bin of calculated QP neutrons.  In Tab. \ref{tab:reconstructed_qpn} we report the mean and widths of those distributions for the $^{70}$Zn+$^{70}$Zn reaction.
For the HIPSE-SIMON simulated data, Eq. \ref{eq.nqp} gives the expected number of neutrons to associate to the QP to within $1$ neutron. This is true for events with between $3$ and $6$ neutrons added back into the QP, which correspond to about $52\%$ of the analysed events. For events with $N_{QP} =0-8$, the calculated $N_{QP}$ value is in agreement with the raw QP-emitted neutron multiplicity within less than $1.5$ neutrons. This corresponds to about $92\%$ of the analysed events.
The expected number of neutrons to associate to the QP is reproduced by Eq.\ref{eq.nqp} for CoMD results to within $1$ for about  $70\%$ of the analysed events. This corresponds to events with $6$ or fewer neutrons added back to the QP. For events with $N_{QP}=0-11$, which correspond to about $92\%$ of the available statistics, the calculated $N_{QP}$ is in agreement with the raw QP-emitted neutron multiplicity within less than $1.5$ neutrons. Therefore, this method is reasonably accurate and sufficiently precise as to allow for the study of well-defined QP sources. Nevertheless, the width of the distribution of the raw number of QP-emitted neutrons for each value of the calculated multiplicity of QP-assigned neutrons introduces an uncertainty into the determination of the QP composition. In any analysis, the impact of such uncertainty on the results should be taken into account.

%

\begin{table}
\centering
\begin{tabular}{|c|c|c|c|c|}
\hline
$N_{QP}$ & $\langle M_{QP_{n}}^{raw} \rangle$ & $\sigma(M_{QP_{n}}^{raw})$& $\langle M_{QP_{n}}^{raw} \rangle$ & $\sigma(M_{QP_{n}}^{raw})$\\
\hline
        & \multicolumn{2}{|c|}{HIPSE} & \multicolumn{2}{|c|}{COMD} \\
          
\hline
$0 $&$ 1.46 $&$ 1.10 $&$ 0.51 $&$ 0.74$\\
$1 $& $2.13 $& $1.21$ & $1.47$ & $1.08$\\
$2 $ & $3.02$ & $1.64$ & $2.69$ &$ 1.34$\\
$3 $& $3.67 $& $1.68$ & $3.76$ & $1.58$\\
$4 $&$ 4.06$ &$ 1.75$ & $4.66$ & $1.78$\\
$5 $& $4.72$ & $1.88 $& $5.88$ & $2.03$\\
$6 $& $5.28$ & $1.97$ & $6.81$ & $2.16$\\
$7 $&$ 5.94$ &$ 2.04$ & $8.44$ & $2.52$\\
$8 $& $6.59$ & $2.14$ & $9.61$ & $2.49$\\
$9 $&  $-$    &  $-$     & $10.63$ & $2.73$\\
$10$&  $-$     &  $-$     & $11.58$ &$ 2.65$\\
$11$&  $-$     &  $-$     & $12.49$ & $2.77$\\
\hline
\end{tabular} 
\caption{Average ($\langle M_{QP_{n}}^{raw} \rangle$) and $\sigma$ of the QP-emitted neutron multiplicity distribution, plotted in Fig.\ref{fig:howgoodOurFormulaIs}, for each value of reconstructed QP neutron multiplicity ($N_{QP}$) for HIPSE-SIMON and CoMD simulated data.}
\label{tab:reconstructed_qpn}
\end{table}

\section{Experimental QP neutron distributions}\label{sec:exp}
\begin{figure}[h!]
\centering
\includegraphics[width=0.95\columnwidth]
{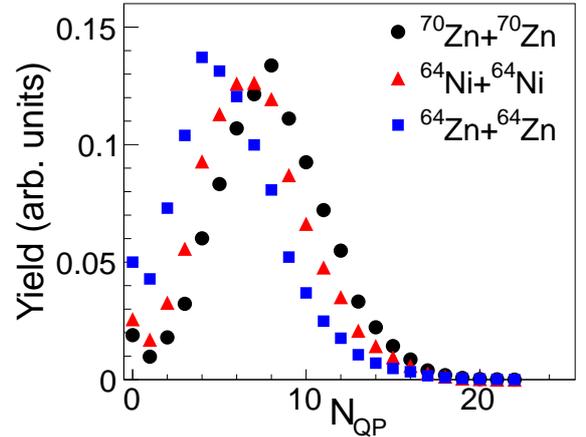}
\caption{(Color online) Reconstructed experimental QP neutron multiplicity as given by Eq.\ref{eq.nqp}, for the $^{70}$Zn+$^{70}$Zn, $^{64}$Ni+$^{64}$Ni and $^{64}$Zn+$^{64}$Zn systems.}
\label{fig:expNdistr}
\end{figure}

As an example, the described criteria for assigning detected free neutrons to the QP fragmenting source was applied to experimental data on $^{70}$Zn+$^{70}$Zn, $^{64}$Ni+$^{64}$Ni and $^{64}$Zn+$^{64}$Zn measured with the $4\pi$ NIMROD-ISiS array described in Sec.\ref{sec:Experimental_setup}. The QP source was reconstructed, event-by-event, applying sequencially the \textit{VCut}, the \textit{SumZ} (the total detected charge was required to be within $25$ and $30$), and the \textit{QCut} to the data. 
To be applied to experimental data, Eq. \ref{eq.nqp} needs to be modified as:
\begin{equation}\label{eq.nqp_exp}
N_{QP} = \frac{N_{det} - N_{background}} {(\varepsilon_{QP}-\frac{N_{T}}{N_{P}}\varepsilon_{QT}) \frac{\varepsilon_{Cf}}{\varepsilon_{Cf_{GEANT}}}}.
\end{equation}
The measured neutron multiplicity ($N_{det}$) is corrected for the measured background multiplicity ($N_{background}$), which is determined using the gating system discussed in Sec. \ref{sec:Experimental_setup}.

The $\varepsilon_{Cf}/\varepsilon_{Cf_{GEANT}}$ term is the ratio of the efficiencies for  a $^{252}$Cf source positioned at the target position  determined during the experimental campaign ($\varepsilon_{Cf}$) and by the GEANT-3 simulation ($\varepsilon_{Cf_{GEANT}}$), described in Sec. \ref{sec:analysis}. Indeed the GEANT-3 simulation, on which the experimental filter is based, predicts a smaller number of neutrons to be detected with respect to the experimental one. This is due to a not perfect reproduction of all the processes involved in the neutron detection in the simulation. Therefore the efficiencies ($\varepsilon_{QP}$ and $\varepsilon_{QT}$) that GEANT-3 calculates are lower than the real values. While this is not important when dealing with simulated data, since both $N_{det}$ and the QP and QT efficiencies are reduced of the same factor, it has to be taken into account in experimental data. This is done by introducing the term $\varepsilon_{Cf}/\varepsilon_{Cf_{GEANT}}$ in Eq. \ref{eq.nqp_exp}.
The experimental 
$\varepsilon_{Cf}$ was measured to be $0.7$, while the GEANT-3 value was $0.6$. As for the QP and QT efficiencies, $\varepsilon_{QP}$ and $\varepsilon_{QT}$, we used the values obtained from the HIPSE-SIMON calculation reported in Tab. \ref{tab:hipse_QPeff_lab_and_Vcut}. Equation \ref{eq.nqp_exp} was applied to the experimental free neutron multiplicities measured by Neutron Ball for the three reactions and allowed the determination of the QP mass.  The QP $N/Z$ distributions are centered around $1.35$, $1.30$ and $1.20$, with a width (RMS) of $0.13$, for the  $^{70}$Zn+$^{70}$Zn, $^{64}$Ni+$^{64}$Ni  and $^{64}$Zn+$^{64}$Zn reactions, respectively, as it was shown in \cite{marini2011}. In Fig. \ref{fig:expNdistr} we present the obtained reconstructed QP neutron multiplicities for the $^{70}$Zn+$^{70}$Zn (circles), $^{64}$Ni+$^{64}$Ni (triangles) and $^{64}$Zn+$^{64}$Zn (squares). The distributions were normalized to $1$. The highest mean $N_{QP}$ value ($\langle N_{QP} \rangle = 7.95$) is obtained for the most neutron-rich reaction ($^{70}$Zn+$^{70}$Zn) and it decreases as the average QP isospin decreases. Indeed $\langle N_{QP} \rangle$ values of $6.69$ and $5.52$ are found for $^{64}$Ni+$^{64}$Ni  and $^{64}$Zn+$^{64}$Zn, respectively. These observations are consistent with the expected enhanced emission of neutrons  for more neutron-rich systems.

\section{Conclusions}\label{sec:conclusions}
The QP reconstruction procedure - in particular the method used to determine the multiplicity of QP-emitted free neutrons - was investigated in detail through two different simulation codes, CoMD and HIPSE-SIMON.

The ability of the velocity cut to select QP-emitted neutrons (and protons) was carefully investigated.   The methods based on the velocity selection (typically used for charged particles) and on efficiencies calculated from QP and QT-tagged neutrons from HIPSE are consistent.
Both CoMD and HIPSE-SIMON simulated data were used to determine the overall and source-specific detection efficiencies of the TAMU Neutron Ball for multifragmentation events. The agreement within $2\%$ of the results obtained from the two codes suggests the efficiency values to be largely model-independent.
The narrow correlation observed between the actual emitted and detected neutron multiplicities indicates that measured free neutron multiplicities can be used to reproduce the actual free neutron multiplicities, even on an event-by-event basis.
The similar narrow correlations observed for predicted and measured QP-emitted neutrons suggest that, with a proper treatment of the efficiencies, the number of neutrons to be assigned to the QP can be determined with an accuracy of around $1.5$ neutrons on average.

Finally, we showed that, with the efficiencies obtained from our calculations, the method using Eq. \ref{eq.nqp} to experimentally determine the QP-emitted free neutron multiplicity is reasonably accurate and sufficiently precise as to allow for the study of well-defined QP sources. Nevertheless, there remains an uncertainty on the QP composition, which should be taken into account by any analysis relying on the QP composition. 
The model independence of our results of the analysed reaction suggests that the described procedure using Eq. \ref{eq.nqp} and the obtained efficiency values can be used in future analysis with different systems and energies provided that efficiencies are recalculated for the different reacting systems.

\section*{Acknowledgements}
The authors would like to thank all their colleagues who have contributed at various stages of this work. We thank R.~Wada  for stimulating discussions.
Part of this work was supported by the U.S. DOE grant DE-FG03-93ER40773 and the
Robert A. Welch Foundation grant A-1266 and one of us (P.M.) also acknowledges the Cyclotron Institute for full financial support during her stay at Texas A$\&$M University. 

\bibliography{bibliogr}
\bibliographystyle{unsrt}
\biboptions{sort&compress}

\end{document}